\documentclass[journal]{IEEEtran}

\ifCLASSINFOpdf
\else
   \usepackage[dvips]{graphicx}
\fi
\usepackage{url}

\usepackage{graphicx}
\usepackage{booktabs}
\usepackage{array,multirow,multicol}
\usepackage{cite}
\usepackage[colorlinks]{hyperref}
\hypersetup{
colorlinks=true,
linkcolor=blue,
filecolor=blue,
citecolor =cyan,
urlcolor=cyan,
}

\begin{document}

\title{Can Layer-wise SSL Features Improve Zero-Shot ASR Performance for Children's Speech?}

\author{Abhijit Sinha, Hemant Kumar Kathania, Sudarsana Reddy Kadiri and Shrikanth Narayanan (\IEEEmembership{IEEE Fellow})
\vspace{-0.5cm}

\thanks{Abhijit Sinha and Hemant Kumar Kathania are with the Department of ECE, NIT Sikkim, India (e-mail:(phec230023 and hemant.ece)@nitsikkim.ac.in). Sudarsana Reddy Kadiri and Shrikanth Narayanan are with the Signal Analysis and Interpretation Lab (SAIL), University of Southern California, USA (e-mail:(skadiri and shri)@usc.edu)}}

\maketitle
\begin{abstract}
Automatic Speech Recognition (ASR) systems often struggle to accurately process children's speech due to its distinct and highly variable acoustic and linguistic characteristics. While recent advancements in self-supervised learning (SSL) models have greatly enhanced the transcription of adult speech, accurately transcribing children's speech remains a significant challenge. This study investigates the effectiveness of layer-wise features extracted from state-of-the-art SSL pre-trained models - specifically, Wav2Vec2, HuBERT, Data2Vec, and WavLM in improving the performance of ASR for children's speech in zero-shot scenarios. A detailed analysis of features extracted from these models was conducted, integrating them into a simplified DNN-based ASR system using the Kaldi toolkit. The analysis identified the most effective layers for enhancing ASR performance on children's speech in a zero-shot scenario, where WSJCAM0 adult speech was used for training and PFSTAR children speech for testing. Experimental results indicated that Layer 22 of the Wav2Vec2 model achieved the lowest Word Error Rate (WER) of 5.15\%, representing a 51.64\% relative improvement over the direct zero-shot decoding using Wav2Vec2 (WER of 10.65\%). Additionally, age group-wise analysis demonstrated consistent performance improvements with increasing age, along with significant gains observed even in younger age groups using the SSL features. Further experiments on the CMU Kids dataset confirmed similar trends, highlighting the generalizability of the proposed approach.
\end{abstract}


\IEEEpeerreviewmaketitle

\section{Introduction}
 \label{sec:sec1}
Automatic Speech Recognition (ASR) technologies have seen substantial progress in recent decades. However, accurately transcribing children's speech remains a significant challenge due to its unique acoustic and linguistic characteristics \cite{koenig2008speech,10.21437/interspeech.2020-3037,yeung2018difficulties}. Children’s speech differs markedly from adult speech, with developmental variations in pronunciation, speaking rate, pitch and evolving vocal tract configurations \cite{lee1999acoustics,gerosa2007acoustic,Vorperian2007VowelAS}. These differences, compounded by the limited availability of annotated children's speech datasets \cite{claus2013survey,feng2024towards,Sukhadia2024ChildrensSR} hinders the development of robust ASR models for Children. The issue is especially evident in zero-shot scenarios, where models trained on adult speech must adapt to the distinct characteristics of children's speech.

To address the limited availability of children's speech data, researchers have explored various techniques for both data augmentation and speech adaptation. Key approaches include time-scale modification \cite{Fan2023UsingMA, sinha2024effect}, formant modification \cite{9746281,kathania_2020, KATHANIA-2022-Speech-comm}, and vocal tract length normalization \cite{Patel2024ImprovingEM}. These methods enhance training datasets and improve model robustness during testing, effectively addressing the challenges posed by limited data in both phases. More advanced techniques, including transfer learning \cite{Rolland2022MultilingualTL, Thienpondt2022TransferLF,shivakumar2022end}, domain adaptation \cite{Fan2022DRAFTAN}, and voice conversion to simulate diverse speech characteristics \cite{shuyang2023data,ANKITA2024104385}, have been proposed to generate richer, more representative datasets. Additionally, text-to-speech synthesis \cite{kadyan2021synthesis,9764693,10256249} has been used to create synthetic data that closely mimics children's speech. These methods aim to provide more diverse and representative training data, improving the accuracy of ASR models for children's speech.

Self-supervised learning (SSL) has recently become a pivotal technique in ASR, enabling models to learn robust speech representations from vast amounts of unlabeled audio data \cite{baevski2020wav2vec, hsu2021hubert, baevski2022data2vec, chen2022wavlm, radford2023robust}. Moreover, research indicates that fine-tuning these pre-trained models on children's speech data significantly enhances ASR performance for this demographic \cite{Fan2022DRAFTAN, fan2022towards, jain2023wav2vec2, Li2024AnalysisOS}. However, fine-tuning generally demands a substantial dataset to achieve optimal results. Given the scarcity of large-scale children's speech corpora, it is critical to explore alternative strategies that maximize the efficiency of the available data to maintain high ASR performance.

In this context, our study aims to improve ASR performance for children's speech, particularly in zero-shot scenarios. We leverage features extracted from state-of-the-art SSL models, including Wav2Vec2 \cite{baevski2020wav2vec}, HuBERT \cite{hsu2021hubert}, Data2Vec \cite{baevski2022data2vec}, and WavLM \cite{chen2022wavlm}, which learn rich, contextualized speech representations from large-scale unlabeled data. The main contribution of our work lies in a systematic, layer-wise analysis of these models to identify which transformer layers most effectively transfer to children's speech. Specifically, we address the following research questions:
\begin{itemize}
\item \textbf{How do SSL models perform in zero-shot ASR for children's speech?}
The goal is to assess the zero-shot capabilities of these models in adapting to children's speech patterns directly from pre-trained features. 
    
\item \textbf{How do features from each layer impact zero-shot ASR performance for children's speech?} 
This analysis focuses on identifying which layers provide the most informative features, to optimize recognition accuracy for children's speech.

\item \textbf{How do these features perform across different age groups in children's speech?} 
By analyzing age-specific performance, we aim to uncover how recognition accuracy varies with age, to inform model adaptation strategies for diverse age cohorts.
\end{itemize}

\section{Proposed Zero-Shot ASR Utilizing Layer-Wise SSL Features}
\label{sec:sec2e}

\begin{figure}[h!]
\centering
\includegraphics[width=\columnwidth]{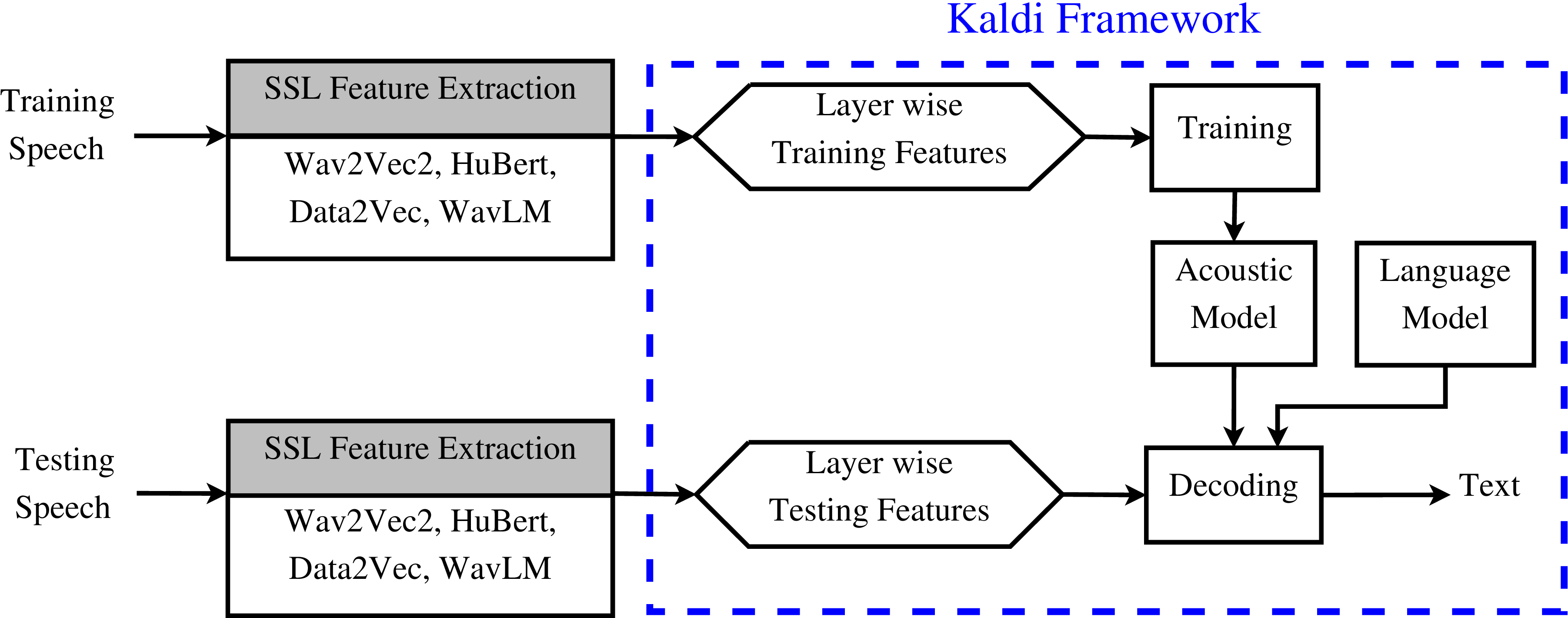}
\caption{The proposed zero-shot ASR framework. The framework integrates SSL models to extract features at different layers, which are then  used as input to the Kaldi ASR system.}
\label{fig:block}
\end{figure}

The proposed framework, illustrated in Figure \ref{fig:block}, presents the architecture for zero-shot ASR on children's speech using features extracted from multiple state-of-the-art SSL models. We leverage four pre-trained SSL models: Wav2Vec2-Large-960h-lv60-self \cite{baevski2020wav2vec}, HuBERT-Large-LS960-ft \cite{hsu2021hubert}, Data2Vec-Audio-Large-960h \cite{baevski2022data2vec}, and WavLM-Large \cite{chen2022wavlm}. These models, which have demonstrated exceptional performance across diverse ASR tasks, generate 1024-dimensional feature representations from the input speech signal. Each SSL model comprises 25 hidden layers, where the first layer (indexed as 0) outputs features from a convolutional neural network (CNN) block, followed by 24 transformer encoder layers (indexed 1 to 24).
The layer-wise features extracted from each model are integrated into a Kaldi-based ASR pipeline \cite{povey2011kaldi}, which trains a deep neural network (DNN) acoustic model \cite{DNN_Hinton_SP_mag}. Note that all SSL models remain frozen: we use only the publicly available pre‐trained checkpoints on Hugging Face (no additional fine‐tuning on either WSJCAM0 or PFSTAR).

\vspace{-10pt}

\vspace{-15pt}
\section{Database and Experimental Setup}
 \label{sec:sec3}

\subsection{Database}
This study employs two widely used British English speech corpora: WSJCAM0 \cite{wsjcam0} for training and PFSTAR \cite{PFStar_database} for testing ASR models. WSJCAM0 is one of the largest spoken corpora of adult British English, containing recordings from 140 speakers, each providing approximately 110 utterances. For this study, the training set from WSJCAM0, consisting of 15.5 hours of data from 92 speakers, was used. 

The PFSTAR children's speech dataset, on the other hand, includes recordings of children aged 4 to 14 years in British English. The PFSTAR training set includes 8.3 hours of recordings from 122 speakers. For testing, a subset of the PFSTAR dataset was used, comprising 1.1 hours of read British English speech data from 60 speakers (28 female, 32 male) aged between 4 and 13 years \cite{KATHANIA-2022-Speech-comm,SHAHNAWAZUDDIN2020213}.

\subsection{Kaldi Framework}

The Kaldi toolkit \cite{povey2011kaldi} was used to build both the baseline and SSL‐enhanced ASR systems. For the baseline, 40‐dimensional MFCCs were extracted (20 ms frames, 10 ms shift) and normalized with fMLLR; the DNN acoustic model \cite{DNN_Hinton_SP_mag} had five hidden layers of 1,024 nodes, trained for 30 epochs (learning rate 0.005, then 0.0005). Decoding employed a bigram language model trained on PFSTAR transcripts (excluding test utterances), following prior zero‐shot children’s ASR work \cite{shahnawazuddin2024effect}. When directly decoding with pre‐trained SSL models, no external LM was used.

\vspace{-0.3cm}

\subsection{Layer-wise SSL Features}
\label{ssec:ExperimentalSetup}

The experiments utilized four state-of-the-art SSL models: Wav2Vec2-Large-960h-lv60-self \cite{baevski2020wav2vec}, HuBERT-Large-LS960-ft \cite{hsu2021hubert}, Data2Vec-Audio-Large-960h \cite{baevski2022data2vec}, and WavLM-Large \cite{chen2022wavlm}, which will be referred to as Wav2Vec2, HuBERT, Data2Vec, and WavLM, respectively, throughout the paper. These models, trained on large-scale unlabeled speech data, are designed to learn robust speech representations, making them highly effective for a wide range of ASR tasks. The Wav2Vec2 model was pre-trained on 60,000 hours of unlabeled data and fine-tuned on 960 hours of labeled data. HuBERT used the same unlabeled data but with a masked prediction strategy. The Data2Vec model is also pre-trained on 60,000 hours of unlabeled data and fine-tuned on 960 hours of labeled data, utilizing a future frame prediction approach to enhance its representation learning capabilities. WavLM was pre-trained on 94,000 hours from various sources and fine-tuned on 960 hours of labeled data. Each model employs CNNs to transform raw speech into latent representations, effectively capturing local acoustic features from the waveform. The features extracted by the CNNs are subsequently input into Transformer encoders, which are designed to capture long-range dependencies. 

Each model consists of 25 (0-24) hidden layers, and for each speech signal, we extracted the outputs from all layers, resulting in 25 distinct feature matrices. Each matrix contains a sequence of feature vectors corresponding to the input speech frames, with each feature vector having a dimension of 1024. These SSL extracted features were then integrated into the Kaldi pipeline, replacing the traditional MFCC features.

\vspace{-0.2cm}
\subsection{Experiments}
In this study, a series of experiments was conducted to evaluate the effectiveness of SSL models for zero-shot ASR on children's speech. The experimental design included the following:

\begin{itemize} 

\item \textbf{Baseline Zero-Shot Performance:} We established baseline ASR performance using MFCC features with Kaldi and by decoding the test set directly with SSL models.

\item \textbf{Layer-wise Feature Performance:} We analyzed the impact of features extracted from different layers of the SSL models on ASR performance, identifying the most effective layers for recognizing children's speech.

\item \textbf{Age Group-wise Analysis:} We examined recognition accuracy across various age groups to evaluate how well SSL models generalized to children's speech at different age groups.

\item \textbf{Comparison with Previous Studies:} Our results were compared with those of prior studies, highlighting the performance gains achieved by our proposed approach that incorporates SSL features into the ASR system.

\end{itemize}

\section{Results and Discussion}
\label{sec:sec4}
Section \ref{sec:Baseline} discusses the baseline zero-shot performance of the models. Section \ref{sec:layer_wise} presents the results from the layer-wise analysis of the SSL models. Section \ref{sec:age_wise} outlines the findings from the age group-wise analysis, while Section \ref{sec:comparision} compares the results of the proposed approach with those from previous studies.

\vspace{-0.3cm}
\subsection{Baseline Zero-Shot Performance}
\label{sec:Baseline}

\begin{figure}[h!]
\centering
\includegraphics[width=\columnwidth]{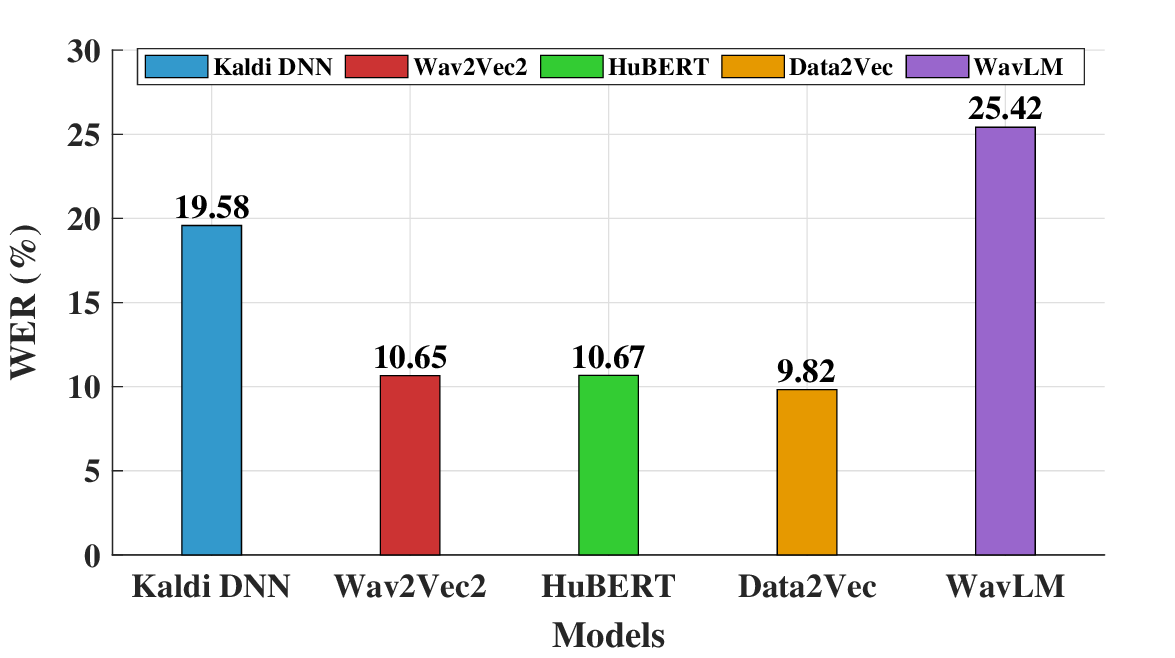}
\vspace{-24pt}
\caption{Comparison of baseline zero-shot WER between the Kaldi DNN model, which utilizes MFCC features, and various SSL models for the PFSTAR dataset.}
\label{fig:baseline}
\end{figure}

This section presents the baseline zero-shot results for the Kaldi DNN ASR model and the SSL models. Figure \ref{fig:baseline} compares the zero-shot WER performance of various SSL models alongside the Kaldi-based ASR system, which employs MFCC features. The results indicate that, all SSL models outperform Kaldi in zero-shot ASR, except WavLM, which may overfit due to additional pretraining objectives like speech enhancement and speaker modeling. Notably, Data2Vec achieves a WER of 9.82\%, followed closely by HuBERT and Wav2Vec2, with WERs of 10.67\% and 10.65\%, respectively. This comparison underscores the superior performance of SSL models in transcribing children's speech without task-specific fine-tuning. The results demonstrate that SSL models generally yield better transcription accuracy in zero-shot settings, highlighting the effectiveness of leveraging pre-trained representations.

The subsequent experiments concentrate on the three best-performing SSL models identified in Figure \ref{fig:baseline}.

\vspace{-13pt}

\vspace{-20pt}

\subsection{Layer-wise Feature Performance}
\label{sec:layer_wise}
\begin{figure}[!h]
   \centering
   \includegraphics[width=\columnwidth, height=4.5cm]{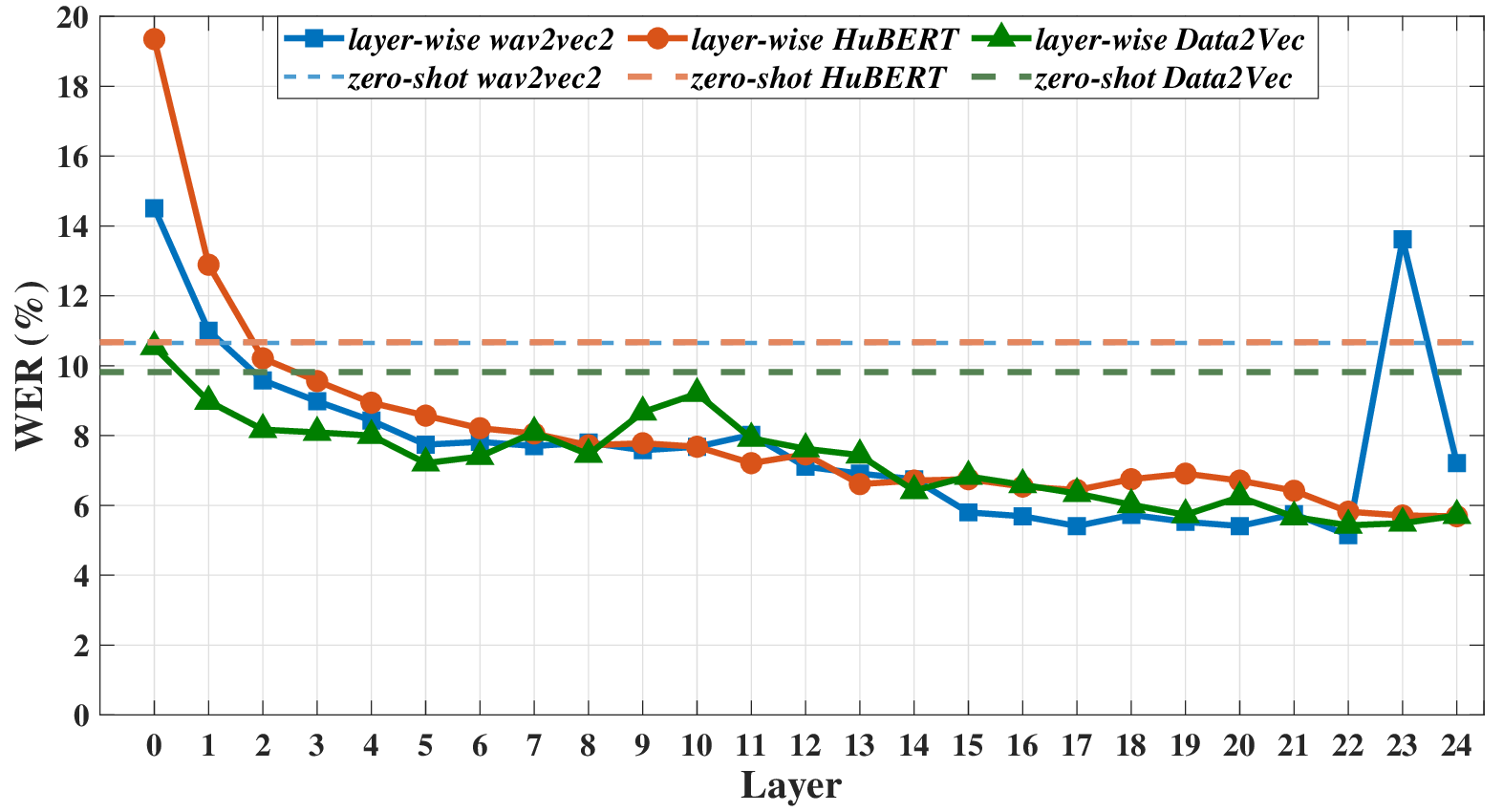}
   \vspace{-0.8cm}
   \caption{ASR performance of the PFSTAR dataset based on layer-wise features extracted from three SSL models: Wav2Vec2, HuBERT, and Data2Vec. The baseline zero-shot WERs are also shown for comparison.}

   \label{fig:figure1}
\end{figure}

This section examines the layer-wise performance of the three selected SSL models: Wav2Vec2, HuBERT, and Data2Vec. 
The analysis aims to identify the optimal layers that yield the best results in zero-shot conditions.
Figure \ref{fig:figure1} provides a comprehensive comparison of layer-wise zero-shot ASR performance for children's speech, utilizing features extracted from each of the 25 (0-24) layers of Wav2Vec2, HuBERT, and Data2Vec. The analysis reveals significant variations in ASR performance across different layers, which can be attributed to the distinct types of features captured at various depths within these models. In the initial layers (0-5), the WER is notably higher for all three models, indicating that these layers predominantly capture low-level acoustic features, which are less effective for speech recognition tasks. As the analysis progresses to the intermediate layers (6-15), a noticeable improvement in WER is observed, reflecting a shift in the models toward capturing more abstract and meaningful features. For instance, WER for Wav2vec2 drops from 7.82\% at layer 6 to 5.80\% at layer 15, similarly, HuBERT and Data2Vec demonstrate comparable improvements, indicating that these layers capture more relevant features for ASR.

\vspace{-10pt}

\begin{table}[!htbp]
    \centering
    \caption{Baseline zero-shot WER (\%) and best performing layers of three SSL models (proposed zero-shot). The relative improvement (Rel. Imp.) indicates WER reduction relative to each model's baseline.}
    \vspace{-8pt}
    \resizebox{8.0cm}{!}{
    \begin{tabular}{ccccc}
    \toprule
        \multirow{2}{*}{\textbf{Model}} & \textbf{Baseline} & \multirow{2}{*}{\textbf{Best Layer}} & \textbf{Proposed} &  \multirow{2}{*}{\textbf{Rel. Imp.(\%)}} \\ 
        & \textbf{Zero-Shot} & & \textbf{Zero-Shot} &\\
    \midrule
        Wav2Vec2  & 10.65  & 22  & \textbf{5.15}  & 51.64 \\
        HuBERT    & 10.67  & 24  & 5.69  & 46.67 \\
        Data2Vec  & 9.82   & 22  & 5.43  & 44.70 \\
    \bottomrule
    \end{tabular}}
    \vspace{-10pt}
    \label{tab:Table3}
\end{table}

The most significant reduction in WER is observed in the later layers (16-24), where the models capture highly abstract and relevant features essential for accurate speech recognition. In these layers, phonemic information appears to be separated from age-specific attributes, allowing the model to work well across different age speakers. Wav2Vec2 achieves its lowest WER of 5.15\% at layer 22 but experiences a sudden spike to 13.62\% at layer 23, suggesting that deeper layers may not always provide the optimal features for recognizing children's speech. In contrast, HuBERT demonstrates a more stable decrease in WER, reaching 5.69\% at layer 24 without such fluctuations, indicating a more consistent feature extraction process. Data2Vec also performs well in the later layers, with a lowest WER of 5.43\% at layer 22. Table \ref{tab:Table3} summarizes the performance evaluation of the best-performing layer features from three SSL models: Wav2Vec2, HuBERT, and Data2Vec. The table includes the zero-shot WER for each baseline model, the best performing layer identified for each (proposed zero-shot), and the relative improvement (Rel. Imp.) percentage, indicating the reduction in WER attained by our method over the baseline SSL models. Notably, our approach demonstrates substantial enhancements in ASR performance, achieving a relative improvement of 51.64\% for Wav2Vec2, 46.67\% for HuBERT, and 44.70\% for Data2Vec. These results illustrate the effectiveness of our methodology in addressing the challenges of recognizing children's speech in a zero-shot context. The WER reduction from the MFCC baseline (19.58\%) to the best SSL layer (5.15\%) is statistically significant, with a 95\% confidence interval.

    \vspace{-10pt}
\begin{figure}[!h]
   \centering
   \includegraphics[width=\columnwidth, height=4.5cm]{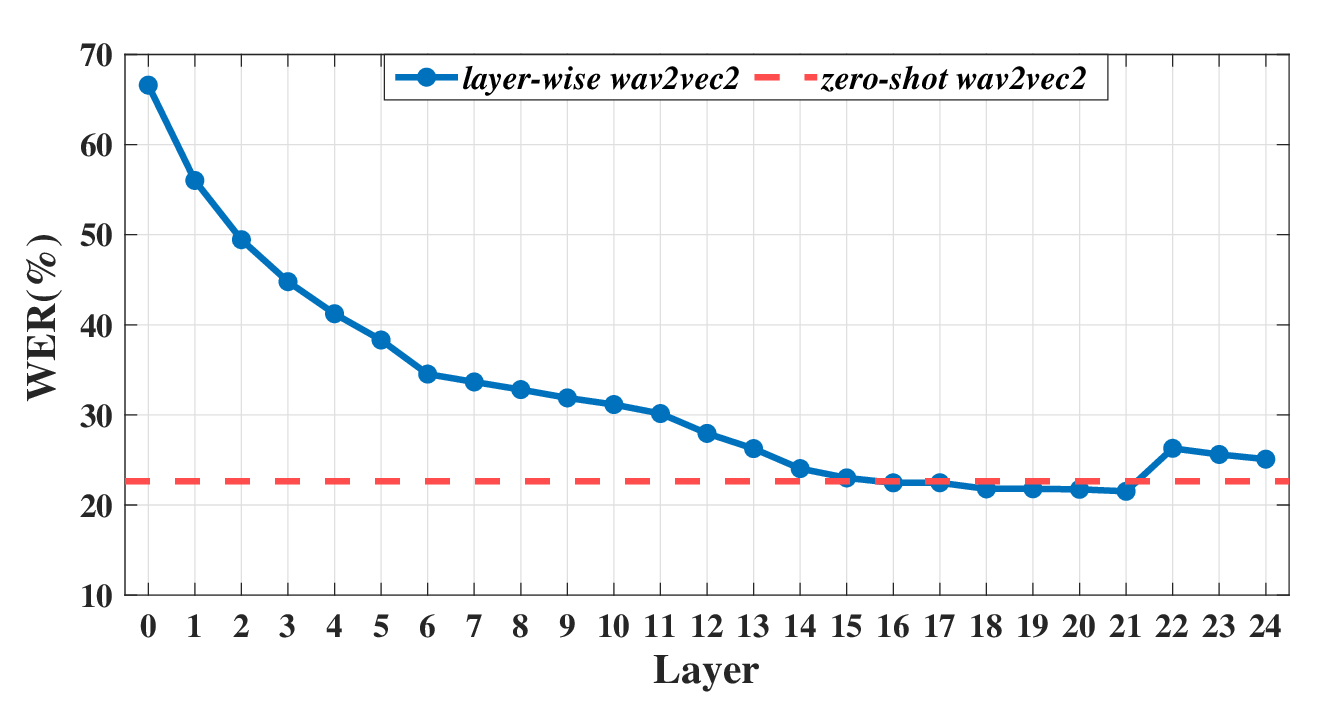}
   \vspace{-25pt}
   \caption{ASR performance of the CMU Kids dataset based on layer-wise features extracted from the Wav2Vec2 model. The baseline zero-shot WER is also shown for comparison.}
    \vspace{-15pt}
   \label{fig:figure2}
\end{figure}

\subsection{Layer‐Wise Generalization on CMU Kids Corpus}
\label{sec:cmukids_layerwise}

To further validate our PFSTAR layer‐wise trends, we applied the best performing SSL model (Wav2Vec2‐large‐960h‐lv60‐self) to a zero‐shot analysis on the CMU Kids Corpus \cite{eskenazi1997cmu}. CMU Kids contains 5,180 read‐speech utterances (9 h) from 78 American children (24 males, 52 females) aged 6–11 years. However, evaluating CMU Kids in a zero‐shot setup posed an accent mismatch: WSJCAM0 (used to train the Kaldi DNN) is British English, while CMU Kids is American English. To mitigate this mismatch, we retrained the Kaldi DNN on MiniLibriSpeech \cite{7178964} (American English adult read speech) and then decoded CMU Kids using that model. Figure \ref{fig:figure2} shows the layer-wise zero-shot ASR performance on CMU Kids dataset. The earliest layers (0–3) mirror MFCC performance (66.60\% at layer 0, 44.79\% at layer 3). Layers 4-12 then reduce WER steadily from 41.23\% down to 27.95\%. The lowest WER (21.52\%) occurs at layer 21; beyond that (layer 22 and above), WER rises again (e.g., 26.29\% at layer 22), indicating over‐specialization. This trend in Fig.~\ref{fig:figure2} matches our PFSTAR layer‐wise findings.

Table~\ref{tab:cmukids_wav2vec2} shows the WERs of baseline zero-shot and the best-performing layer 21 of the Wav2Vec2 model for the CMU Kids dataset, showing a 4.89\% relative improvement.

\vspace{-10pt}

\begin{table}[!htbp]
    \centering
    \caption{Baseline zero-shot WER (\%) and best performing layer of the Wav2Vec2 model (proposed zero-shot). The relative improvement (Rel. Imp.) indicates WER reduction relative to baseline.}
    \vspace{-8pt}
    \resizebox{8.0cm}{!}{
    \begin{tabular}{ccccc}
    \toprule
        \multirow{2}{*}{\textbf{Model}} & \textbf{Baseline} & \multirow{2}{*}{\textbf{Best Layer}} & \textbf{Proposed} &  \multirow{2}{*}{\textbf{Rel. Imp.(\%)}} \\ 
        & \textbf{Zero-Shot} & & \textbf{Zero-Shot} &\\
    \midrule
        Wav2Vec2       & 22.63             & 21                  & 21.52             & 4.89                   \\
    \bottomrule
    \end{tabular}}
    \vspace{-7pt}
    \label{tab:cmukids_wav2vec2}
\end{table}

To confirm the robustness of these SSL features with larger training datasets, we further evaluated the two best Wav2Vec2 layers 20 and 21 on the LibriSpeech 100-hour subset. Table~\ref{tab:scaleup} shows the WERs remain consistent across MiniLibriSpeech and LibriSpeech training data, with layer 20 achieving 21.74\% and 21.75\%, and layer 21 achieving 21.52\% and 21.47\%, respectively. These results demonstrate that our findings remain consistent even when using more training data.

\begin{table}[!h]
  \centering
  \vspace{-10pt}
  \caption{WER (\%) for MFCC baseline and best two Wav2Vec2 layers on MiniLibriSpeech vs.\ LibriSpeech (100 h) for CMU Kids dataset.}
  \vspace{-8pt}
  \label{tab:scaleup}
      \resizebox{8cm}{!}{ 
  \begin{tabular}{lcc}
    \toprule
    \multirow{2}{*}{\textbf{Feature / Layer}} 
      & \multicolumn{2}{c}{\textbf{Training Data}} \\ 
    \cmidrule{2-3}
      & \textbf{MiniLibriSpeech} & \textbf{LibriSpeech (100 h)} \\
    \midrule
    MFCC Baseline        & 66.29 & 50.93 \\
    Wav2Vec2 Layer 20    & 21.74 & 21.75 \\ 
    Wav2Vec2 Layer 21    & 21.52 & 21.47 \\ \bottomrule
  \end{tabular}}
  \vspace{-10pt}
\end{table}

\vspace{-10pt}
\subsection{Age Group‐Wise Analysis}
\label{sec:age_wise}

Using the best Wav2Vec2 layers (PFSTAR: layer 22; CMU Kids: layer 21), we compared zero‐shot WER across age groups for both datasets (Table \ref{tab:wer_comparison}). On PFSTAR, the youngest group (ages 4-6) drops from 27.35\% to 13.51\%, while the oldest group (ages 10-13) falls from 7.09\% to 4.09\%. The middle group (ages 7-9) sees an intermediate gain: 8.39\% → 3.75\%. Thus, although 10-13 year olds achieve the lowest absolute WER, the largest absolute improvement occurs for the youngest speakers (4-6 years), indicating that SSL features help most where age‐related variability is greatest.

As per previous results we used MiniLibriSpeech model for age-wise evaluation of CMU Kids dataset. The results shows a WER reduction from 24.58\% to 23.57\% for ages 6-8, and from 17.77\% to 16.69\% for ages 9-11.  Older children (9-11 years) attain a lower absolute WER, but the relative gain is similar across both groups.

\vspace{-10pt}

\begin{table}[!htbp]
    \centering
    \caption{Age group‐wise WER (\%) for PFSTAR and CMU Kids using Wav2Vec2 (zero‐shot).}
    \vspace{-8pt}
    \resizebox{8cm}{!}{ 
    \begin{tabular}{ccccc}
    \toprule
        \multirow{2}{*}{\textbf{Dataset}} & \multirow{2}{*}{\textbf{Age Group}} & \textbf{Baseline} & \textbf{Proposed} & \multirow{2}{*}{\textbf{Rel. Imp.(\%)}}  \\ 
        & &\textbf{Zero‐Shot} & \textbf{Zero‐Shot} &\\
    \midrule
        \multirow{3}{*}{\textbf{PFSTAR}} & Age 4-6   & 27.35 & 13.51 & 50.61 \\ 
          & Age 7-9   & 8.39  & 3.75  & 55.32 \\ 
           & Age 10-13 & 7.09  & 4.09           & 42.30 \\ 
    \midrule
        \multirow{2}{*}{\textbf{CMU Kids}} & Age 6-8   & 24.58 & 23.57 & 4.11 \\ 
          & Age 9-11  & 17.77 & 16.69 & 6.08 \\ 
    \bottomrule
    \end{tabular}
    }
    \label{tab:wer_comparison}
\end{table}

\vspace{-15pt}
\subsection{Comparison with Previous Studies}
\label{sec:comparision}

This section conducts a comparative analysis of our proposed framework against prior studies investigating zero-shot ASR for children's speech, utilizing the PFSTAR dataset as the evaluation benchmark. Table \ref{tab:compared_work} details the performance results from several previous approaches alongside our findings. Notably, our method achieves a WER of 5.15\%, surpassing the performance of earlier methodologies, such as pitch robust BS-MFCC features \cite{shahnawazuddin2021enhancing,shahnawazuddin2024effect} and formant modification  \cite{KATHANIA-2022-Speech-comm} techniques.

\vspace{-10pt}

\begin{table}[!ht]
    \centering
    \small
    \caption{This table compares the performance of our proposed framework with previous studies for zero-shot children asr on the PFSTAR dataset.}
    \label{tab:compared_work}
     \vspace{-8pt}
    \resizebox{8.5cm}{!}{ 
    \begin{tabular}
    {p{1.9cm} p{5cm} p{1cm} p{0.8cm}} 
    \toprule
       \textbf{Author} & \textbf{Methodology }  & \textbf{System}  & \textbf{WER(\%)}  \\ \toprule
        Shahnawazuddin et al \cite{shahnawazuddin2021enhancing}. & Pitch robust BS-MFCC features & TDNN & 9.5 \\ \hline
        
        Kathania et al. \cite{KATHANIA-2022-Speech-comm} & Formant Modification to minimize mismatch between Adult and Child speech & TDNN & 8.69 \\  \hline
         

         Ankita et al. \cite{shahnawazuddin2024effect} & Combined jitter and strength of excitation with MFCC features & TDNN & 7.1 \\ \hline

         Proposed & Layer-wise SSL features. & DNN & \textbf{5.15} \\
           \bottomrule
           
    \end{tabular}}
\end{table}

\vspace{-10pt}
\section{Conclusion}
 \label{sec:sec6} 

This study demonstrates the effectiveness of using layer-wise features from SSL models in a zero-shot ASR system for children's speech. By removing the need for fine-tuning, our approach addresses the data scarcity challenge in child-specific ASR. It outperforms both prior zero-shot systems and standard SSL-based decoding, highlighting the robustness of SSL features even without task adaptation. Layer-wise analysis shows that later layers (16–24) yield better performance, likely due to their ability to capture more abstract, task-relevant representations. Age-wise trends reveal decreasing WER with increasing age, as older children’s speech resembles adult speech more closely. Still, the system performs competitively even for younger age groups, demonstrating strong generalization across diverse speech characteristics.

\bibliographystyle{IEEEbib}
\bibliography{mybib,refs}

\end{document}